\documentclass[showpacs,preprint,amsmath,amssymb, nofootinbib]{revtex4-1}
\newcommand{\be}{\begin{equation}}
\newcommand{\en}{\end{equation}}
\newcommand{\bea}{\begin{eqnarray}}
\newcommand{\ena}{\end{eqnarray}}
\begin{document}
\title{On Ho\v{r}ava-Lifshitz Cosmology}
\author{ Samuel Lepe and Joel Saavedra}
\address{Instituto de F\'{\i}sica, Pontificia
Universidad Cat\'olica de Valpara\'{\i}so, Casilla 4950,
Valpara\'{\i}so, Chile.}
\date{\today}
\begin{abstract}
 We discuss some aspects of the Horava-Lifshitz cosmology with
different matter components considered as dominants at different stages of
the cosmic evolution (each stage is represented by an equation of state
pressure/density=constant). We compare cosmological solutions from this
theory with their counterparts of General Relativity (Friedmann cosmology).
At early times, the Horava- Lifshitz cosmology contains a
curvature-dependent dominant term which is stiff matter-reminiscent and this
fact motivates to discuss, in some detail, this term beside the usual stiff
matter component (pressure=density) if we are thinking in the role that this
fluid could have played early in the framework of the holographic cosmology. Nevertheless, we show that an early stiff matter component is of little
relevance in Horava-Lifshitz cosmology.
\end{abstract}
\maketitle
\section{Introduction}
Searching for a quantum theory of gravity has been a fruitful field for theoretical physics in the last century. In this sense, Ho\v{r}ava proposed in Ref. \cite{Horava:2008jf} a new quantizable theory of gravity, using the ideas from solid state physics. This theory was originally called Ho\v{r}ava-Lifshitz quantum gravity, that is a power-counting renormalizable theory with consistent ultraviolet (UV)  behavior and, on the other hand, the  theory has one fixed point in the infrared limit (IR) namely General Relativity (GR) \cite{Horava:2008jf, Horava:2008ih, Horava:2009if, Horava:2009uw}. Thus, this theory leads to a modification of the Einstein's general relativity at  high energies producing interesting features in cosmology.   The first ideas about this subject have been presented in Refs. \cite{Calcagni:2009ar, Kiritsis:2009sh, Takahashi:2009wc} where the cosmological consequences of Ho\v{r}ava-Lifshitz (HL)  cosmology were studied in a vast detail,  see also Refs. \cite{Nojiri:2010wj, Clifton:2011jh, Brandenberger:2009yt, Mukohyama:2009gg, Cai:2009dx, Saridakis:2009bv, Mukohyama:2009zs, Mukohyama:2009mz, Cai:2009in, Wang:2009azb, Leon:2009rc, Minamitsuji:2009ii, Carloni:2009jc, Gao:2009wn, Kobayashi:2010eh, Maeda:2010ke, Saridakis:2011pk, Bertolami:2011ka}. The new theory has virtues and defects, and its basis  ideas are not free of controversy \cite{Henneaux:2009zb}. In this article we would like to address  an interesting point concerning HL cosmology that is still an open problem. In particular, we are seeing a behavior of HL cosmology for different kinds of matter content of the universe such as dust, cosmological constant and phantom matter. The present paper is organized as follows. In Sec. I we discuss the HL cosmology for different kinds of the matter and also we present general settings and constrains in the theory. In Sec. II we analyze the role of stiff-like matter taken as the dominant at early times. Finally, in Sec. III we give some final remarks.

\section{Ho\v{r}ava-Lifshitz Cosmology}
In this section we would like to review the main results of Hl cosmology and its implications for different kind of mater. Also, we are doing the comparison between the behaviors of HL and GR cosmology.
Let us first to introduce a cosmological model under
consideration based on Ref. \cite{Mukohyama:2010xz}.
It is well-known that in Ho\v{r}ava-Lifshitz gravity we do not have the
Hamiltonian constraint and therefore, from the formal cosmological point of
view, also we do not have one Friedmann equation as in standard cosmology. In this theory the starting point is the dynamical equation for flat Friedmann-Robertson-Walker (FRW) spacetime 
\begin{equation}
\eta \left( 2\dot{H}+3H^{2}\right) =-p,  \label{eq.1}
\end{equation}%
where we introduced a parameter
\begin{equation}
\eta =\frac{1}{2}\left( 3\lambda -1\right) ,  \label{eq.2}
\end{equation}%
and $\lambda $ represents a dimensionless constant. The value of $\lambda$ is fixed by the diffeomorsphism invariance of four-dimensional general relativity. At this point, we would like to emphasize that in HL cosmology any value of the constant $\lambda$ is consistent with a foliation that preserves the diffeomorphism invariance. For example,  it was shown in Ref. \cite{Gumrukcuoglu:2011xg} that for $1/3<\lambda
<1$ (or equivalently $0<\eta <1$) the scalar graviton is a ghost and  is
ghosts-free if $\lambda <1/3$ (that is $\eta <0$) and  $\lambda
>1$  (or $\eta >1$). 
The field equation for matter field is described satisfies a
non-conservation equation, and at high energy limit, or early times limit, we
can write it in the following form 
\begin{equation}
\dot{\rho}+3H(\rho +P)=-Q,  \label{eq.3}
\end{equation}
where $Q$ represents the rate of energy non-conservation, and  the low energy limit  can be recovered only if $Q\rightarrow
0 $. We can conclude that the energy non-conservation
 is an effect of high-energy physics. Now,  equations (\ref{eq.1})
and (\ref{eq.3}) contain enough information  to describe a
cosmological evolution of the spacetime, together with the equation
of state (EoS) $p=\omega \rho $. Indeed, using this dynamical equation plus the
non-conservation equation we can obtain one first integral in the form
\begin{equation}
3H^{2}=\frac{1}{\eta }\rho +\frac{C\left( t\right) }{a^{3}},  \label{eq.4}
\end{equation}%
and this result  can be interpreted as a Friedmann equation in HL gravity where
the integration "constant" is given by 
\begin{equation}
C\left( t\right) =C_{0}+\frac{1}{\eta }\int_{t_{0}}^{t}d\tau a^{3}\left(
\tau \right) Q\left( \tau \right) .  \label{eq.c}
\end{equation}%
In order to obtain a cosmological description from this kind of theories, we
would like to discuss first some well known results from GR in the context of HL
cosmology. The subscript $0$ indicates quantities today ($t_{0}$).

 \textbf{ $\bullet$ \,Dust matter:\ $p=0\longleftrightarrow \omega =0$
} \vspace{0.5cm} 

 In this case from (\ref{eq.1})
we find exactly the same solution for the Hubble parameter as in GR, i.e., 
\begin{equation}
H\left( t\right) =H_{0}\left[ 1+\frac{3}{2}H_{0}\left( t-t_{0}\right) \right]
^{-1},  \label{eq.5}
\end{equation}%
and for the cosmic scale factor reads
\begin{equation}
a\left( t\right) =a_{0}\left[ 1+\frac{3}{2}H_{0}\left( t-t_{0}\right) \right]
^{2/3}.  \label{eq.6}
\end{equation}%
It is well known in GR, when we have $\omega =0$,  an evolution is described
by an energy density proportional to $a^{-3}$ ( this result is obtained from the conservation energy density equation $\dot{\rho}+3H(\rho +P)=0$)
and the formal solution of the cosmological evolution in GR is
identical to Eqs. (\ref{eq.5}) and (\ref{eq.6}), obtained for HL cosmology.  Although the formal solutions for the Hubble parameter and the scale factor from HL and GR are identical in this case, the dynamical behavior of the energy density is absolutely different, because, in
HL cosmology the energy density is not fixed by the non-conservation
equation  (\ref{eq.3}). Finally for the dust matter case, we would like to mention that (\ref%
{eq.5}) and (\ref{eq.6}) are also solutions of (\ref{eq.1}) if we set $\rho =0$
and $\eta $ is finite. Thus,  these equations can also be seen as a
self-decelerated evolution. 

In a nutshell, both $\omega =0$ and $\rho =0$  lead
 to the same solutions described by Eqs. (\ref{eq.5}) and (\ref{eq.6}). Therefore, in  case of dust matter we  found that the effect of high energy
physics is reflected in the fact that during the early stages, the rate of energy non-conservation is not constant  $Q(t)\neq 0$,
and we have an interchange of energy the between the mater  of the
universe and some unknown source.

 \textbf{ $\bullet$ \,Cosmological constant: \textit{$\omega =-1$. }
} \vspace{0.5cm} 

In this case, the conservation
law 

\begin{equation}
\dot{\rho}=-Q, \label{constant1}
\end{equation}
therefore the behavior of cosmological constant at early universe would be different that the respective one in GR. Then, the interchange of energy between the matter ( described in this case for the cosmological constant kind) and the unknown source gives a non constant behavior for the energy density for our cosmological constant type. We note that  the GR  limit is  recovered if we take  $Q\left( a\rightarrow \infty \right)
\rightarrow 0$, i.e., $\rho =const.$. Then the "role" of the
cosmological constant in HL gravity is different with the respective one in GR.

 \textbf{ $\bullet$ \,Phantom Evolution.
} \vspace{0.5cm}

 In this case, from (\ref{eq.1}) we
can write the following formal solution for the Hubble parameter 
\begin{equation}
H^{2}\left( a\right) =\frac{1}{a^{3}}\left( a_{0}^{3}H_{0}^{2}-\frac{\omega 
}{\eta }\int_{a_{0}}^{a}daa^{2}\rho \left( a\right) \right) ,  \label{eq.16s}
\end{equation}%
and if we take one phantom Ansatz for the energy density given by $\rho \left( a\right) =\rho _{0}\left( a/a_{0}\right)
^{\beta }$ with  $\beta >1$, we get  the explicit form of the formal solution 
\begin{equation}
H^{2}\left( a\right) =H_{0}^{2}\left( \frac{a_{0}}{a}\right) ^{3}\left[ 1-%
\frac{\omega \rho _{0}/3H_{0}^{2}}{\eta \left( 1+\beta /3\right) }\left[
\left( \frac{a}{a_{0}}\right) ^{\beta +3}-1\right] \right] ,  \label{eq.18s}
\end{equation}%
and
the future behavior of the Hubble parameter reads as follow
\begin{equation}
H^{2}\left( a\rightarrow \infty \right) =-\frac{\omega \rho _{0}/3}{\eta
\left( 1+\beta /3\right) }\left( \frac{a_{0}}{a}\right) ^{\beta }.
\end{equation}
One realistic model implies the following restriction for the barotropic index, $\omega <0$. If we use the following setting 
\begin{equation}
\left\vert \omega \right\vert \rho _{0}/3H_{0}^{2}=\left( 1+\beta /3\right)
\eta ,  \label{eq.19s}
\end{equation}%
we obtain one phantom solution for the scale factor given by 
\begin{equation}
a\left( t\right) =a_{0}\left( \frac{2}{\beta H_{0}}\right) ^{2/\beta }\left(
t_{s}-t\right) ^{-2/\beta }\text{ \ \ }and\text{ \ \ }t_{s}=t_{0}+\frac{2}{%
\beta H_{0}}.  \label{eq.20s}
\end{equation}%
It is worthwhile noticing that the solution (\ref{eq.20s}) has a new region for
the phantom evolution for the barotropic index $\omega <0$ instead of the
usual, more restrictive one $\omega <-1$ that appears in GR. 

We  also notice that if $\eta
\rightarrow \infty$, then $\beta \rightarrow -3$ and we can write
\begin{equation}
H^{2}\left( a\right) _{\eta \rightarrow \infty }\rightarrow H_{0}^{2}\left( 
\frac{a_{0}}{a}\right) ^{3}, 
\end{equation}
so that the phantom approach has no sense. According to 
\begin{equation}
\eta =\left\vert \omega \right\vert \left( \rho _{0}/3H_{0}^{2}\right)
\left( 1+\beta /3\right) ^{-1}>1, \label{lol}
\end{equation}
or
\begin{equation}
\left\vert \omega \right\vert \left( \rho _{0}/3H_{0}^{2}\right) >1+\beta
/3, 
\end{equation}
we conclude that the observational data do not allow fulfill the last inequality. On the other hand, if $%
0<\eta <1$ (ghost scalar graviton) we have $\left\vert \omega \right\vert
\left( \rho _{0}/3H_{0}^{2}\right) <1+\beta /3$ and in this case we can have
a phantom phase provided  we can accept an existence of a ghost scalar graviton.

\textbf{ $\bullet$ \,General Settings.
} \vspace{0.5cm} 
We are now interested in  the limit  where the matter sector is decoupled from the gravity sector, where dark matter as an integration constant dominated the evolution of the universe, that its $\eta \rightarrow \infty \left( \lambda
\rightarrow \infty \right) $ \cite{Gumrukcuoglu:2011xg}, in this limit  we can
distinguish several situations. For instance, if $p$ is finite and $\eta
\rightarrow \infty $, from (\ref{eq.1}) we obtain exactly the same solution  given by (\ref{eq.5})
and (\ref{eq.6}) and the Hubble parameter becomes in this limit
\begin{equation}
3H^{2}=\frac{C_{0}}{a^{3}}, \label{heta} 
\end{equation}
thus, a dust-like evolution but $p$  and the barotropic index $\omega$ do not vanish. Also, we would like to notice that if $t\rightarrow \infty $ and $\eta 
$ is finite then
\begin{equation}
3H^{2}\left( t\rightarrow \infty \right) \rightarrow \frac{1}{\eta }\rho
\left( t\rightarrow \infty \right) +\frac{1}{a^{3}\left( t\rightarrow \infty
\right) }\left[ C_{0}+\frac{1}{\eta }\int_{t_{0}}^{\infty }d\tau a^{3}\left(
\tau \right) Q\left( \tau \right) \right] =\frac{1}{\eta }\rho \left(
t\rightarrow \infty \right) , \label{hgrand}
\end{equation}
if at least $Q\left( t\right) $ decreases as $a^{-4}\left( t\right) $ when $%
a\left( t\rightarrow \infty \right) \rightarrow \infty $.  This condition on $Q$ is consistent with the recovering of local invariance for the matter sector, where it is demanded that  $Q\left( a\rightarrow
\infty \right) \rightarrow 0$ and $C\left( t\rightarrow \infty \right)
\rightarrow C_{0}$ (see \cite{Gumrukcuoglu:2011xg}). So, we have the following relation between observational parameters
\begin{equation}
3H_{0}^{2}=\frac{1}{\eta }\rho _{0}+\frac{C_{0}}{a_{0}^{3}}.
\end{equation}

On the other hand, from  (\ref{eq.1}) and (\ref{eq.4}) we can write
\begin{equation}
\dot{H}=-\frac{1}{2}\frac{C_{0}}{a^{3}}-\frac{1}{2\eta }\left[ \left(
1+\omega \right) \rho +\frac{1}{a^{3}}\int_{t_{0}}^{t}d\tau a^{3}\left( \tau
\right) Q\left( \tau \right) \right] , \label{hdot}
\end{equation}
so that if we take the limit $\eta \rightarrow \infty $, and keep finite second term in (\ref{hdot}), we obtain

\begin{equation}
H\left( a\right) =H_{0}%
\sqrt{1+\frac{C_{0}a_{0}^{-3}}{3H_{0}^{2}}\left[ \left( \frac{a_{0}}{a}%
\right) ^{3}-1\right] }. \label{hdot2}
\end{equation}
Furthermore, if $C_{0}a_{0}^{-3}/3H_{0}^{2}<1$, we find

\begin{equation}
H\left( a\rightarrow \infty \right) \rightarrow H_{0}\sqrt{1-\frac{%
C_{0}a_{0}^{-3}}{3H_{0}^{2}}}<H_{0}. \label{hlimit}
\end{equation}
To conclude, in  case with  $\eta \rightarrow \infty $ we find a de Sitter phase at late times. We also notice that this behavior can also be  obtained  if
we choose $\omega =-1$ but keep $\eta $ finite, that is,
\begin{equation}
\dot{H}=-\frac{1}{2}\frac{C\left( t\right) }{a^{3}}\rightarrow \dot{H}\left(
t\rightarrow \infty \right) \rightarrow 0, 
\end{equation}
given that in this limit $C\left( t\rightarrow \infty \right) \rightarrow
C_{0}$. In another words, taking either  $\eta \rightarrow \infty $, for all $t$, and or $t\rightarrow \infty 
$ and $\eta $ finite, we arrive to a de Sitter phase at late times.

 To complete this point we would like to  comment about the acceleration of universe. By using the expression for $\dot{H}$, as well as (\ref{eq.4}), we can
write
\begin{equation}
\dot{H}+H^{2}=-\frac{1}{6\eta }\left[ \left( 1+3\omega \right) \rho +\eta 
\frac{C\left( t\right) }{a^{3}}\right] , 
\end{equation}
and it is straightforward to check that $\eta =1$ plus $C\left( t\right) =0$ implies the
standard expression for the acceleration in GR. Also, if $\dot{H}%
+H^{2}>0$ we must have $\omega <-1/3$ , to avoid violating  the weak
energy condition (WEC) $\rho >0$. Then, we can have quintessence, cosmological
constant or phantom schemes. Now, if $\dot{H}+H^{2}<0$ then $\omega >-1/3$  the WEC is fulfilled, then we have an evolution driven by dark matter.
Finally when $\dot{H}+H^{2}=0$ we obtain the solution
\begin{equation}
H\left( t\right) =H_{0}\left[ 1+H_{0}\left( t-t_{0}\right) \right] ^{-1}, 
\end{equation}
i.e., $\omega =-1/3$ (string gas) as in GR. As a curiosity we notice that
\begin{equation}
\left( \dot{H}+H^{2}\right) _{\omega =-1/3}=\frac{1}{3}\left( \dot{H}\right)
_{\omega =-1}. 
\end{equation}
Finally, by setting  $\dot{H}+H^{2}=0$, we can write the following
expression for the density of energy
\begin{equation}
\rho \left( t\right) =\rho _{0}\left( a/a_{0}\right) ^{-3}+\frac{1}{\eta
\left\vert 1+3\omega \right\vert }\frac{1}{a^{3}}\int_{t_{0}}^{t}d\tau
a^{3}\left( \tau \right) Q\left( \tau \right) , 
\end{equation}
where $C_{0}=\left\vert 1+3\omega \right\vert \rho _{0}a_{0}^{-3}$ and $%
\omega <-1/3$, and we find that $\rho \left( t\rightarrow \infty \right)
\rightarrow \rho _{0}\left( a/a_{0}\right) ^{-3}$, i. e., a dust-like
behavior at late times is driven by a $\omega $ that is dark energy-like.

\section{Stiff-like matter as the dominant term at early times} 
When the higher curvature terms are included in the cosmological evolution the dynamic equation for the Hubble parameter is given by \cite{Mukohyama:2010xz},
\begin{equation}
\eta \left( 2\dot{H}+3H^{2}\right) =-\omega \rho +\frac{\alpha k^{3}}{a^{6}} +\frac{\alpha' k^{2}}{a^{4}}-\frac{k}{a^{2}}+ \Lambda,  \label{eq234}
\end{equation}
where $\alpha$ and $\alpha'$ are constants and $k$ is the spatial
curvature. If we consider the early time limit, the dominant contribution is the term of the form  $\sim a^{-6}$, which could be interpreted as stiff matter like, and also would be explain by the supposition that stiff matter could be one important matter content at the very early universe  \cite{Banks:2004eb, Banks:2008ep}. 
We write (\ref{eq234}) in the form
\begin{equation}
\eta \left( 2\dot{H}+3H^{2}\right) =-\omega \rho +\frac{\alpha k^{3}}{a^{6}}%
\left[ 1+\frac{a^{2}}{\alpha }\left( \alpha `k-a^{2}\right) \right] , 
\label{eq234b}
\end{equation}%
and the incorporation of $\Lambda $ will be discussed later. 
Therefore, we have 
 a new equation to handle that is 
\begin{equation}
3\eta H^{2}=\rho -\frac{\alpha k^{3}}{a^{6}}.  \label{eq.32ss}
\end{equation}
Similarly as before, the parameter $\eta $ can  take any value. The dynamical equation then can be written as
\begin{equation}
\eta \left( 2\dot{H}+3H^{2}\right) =-\omega \rho +\frac{\alpha k^{3}}{a^{6}},
\label{eq.33ss}
\end{equation}%
The aproximation at early times is justified if the scale factor is kept
around $a^{2}\sim \left\vert \alpha `\right\vert $, if $k=1$ and for $k=-1$
we have a more restrictive condition: $a^{2}\left( \alpha `+a^{2}\right)
<<\alpha $ . From Eqs. (\ref{eq.32ss}) and (\ref{eq.33ss}) we obtain the
conservation law (the term proportional to $C\left( t\right) /a^{3}$,
i.e., proportional to $Q$, is not present here given that only we are
holding the dominant contribution $\sim a^{-6}$)
\begin{equation}
\dot{\rho}+3H\left( 1+\omega \right) \rho =0,  \label{eq.34ss}
\end{equation}%
as well as the equation 
\begin{equation}
\dot{H}+3H^{2}=\frac{1}{2\eta }\left( 1-\omega \right) \rho,
\label{eq.35ss}
\end{equation}%
and from these last equations we note that if we do $\omega =1$, we obtain
the usual solution as in GR, i.e. $\rho \sim a^{-6}$ and. We find the
following solution for the Hubble parameter and the scale factor,
respectively, 
\begin{equation}
H\left( t\right) =H_{0}\left[ 1+3H_{0}\left( t-t_{0}\right) \right] ^{-1},
\label{eq.36ss}
\end{equation}%
and 
\begin{equation}
a\left( t\right) =a_{0}\left[ 1+3H_{0}\left( t-t_{0}\right) \right] ^{1/3}.
\label{eq.37ss}
\end{equation}

In order to see the implications of this result, we are comparing it to standard
GR, and the equations $3H^{2}=\rho $ and $\dot{\rho}+6H\rho =0$ ($\omega =1$)
that lead (\ref{eq.36ss}) and (\ref{eq.37ss}). Thus replacing (\ref{eq.36ss})
and (\ref{eq.37ss}) in (\ref{eq.32ss}), we obtain for $\omega =1$ 
\begin{equation}
3\eta H^{2}=\left( \rho _{0}-\frac{\alpha k^{3}}{a_{0}^{6}}\right) \left( 
\frac{a_{0}}{a}\right) ^{6},   \label{eq.38ss}
\end{equation}%
and for $k=1$ we must to have $\rho _{0}a_{0}^{6}>\alpha $ and $\eta >0$ (do
not forget that if $0<\eta <1$ we have a ghost scalar graviton and If $\eta
>1$ we are ghosts-free). If $\rho _{0}a_{0}^{6}<\alpha $, then $\eta <0$ and
we are ghosts-free. For $k=-1$ we must have $\eta >0$.
We want now to discuss an early universe that contains a mixture of stiff
matter-like and cosmological constant,
\begin{equation}
3\eta H^{2}=\rho -\frac{\alpha k^{3}}{a^{6}}+\Lambda ,    \label{s0}
\end{equation}%
and
\begin{equation}
\eta \left( 2\dot{H}+3H^{2}\right) =-\omega \rho +\frac{\alpha k^{3}}{a^{6}}%
+\Lambda .  \label{s1}
\end{equation}%
Then we obtain the $k$-independent equation
\begin{equation}
\dot{H}+3H^{2}=\frac{1}{2\eta }\left[ \left( 1-\omega \right) \rho +2\Lambda %
\right] ,    \label{s2}
\end{equation}%
and for $\omega =1$ and $\eta >0$, the formal solution is
\begin{equation}
H\left( t\right) =\sqrt{\Lambda /3\eta }\left( \frac{1+\Delta _{0}\exp \left[
-2\sqrt{3\Lambda /\eta }\left( t-t_{0}\right) \right] }{1-\Delta _{0}\exp %
\left[ -2\sqrt{3\Lambda /\eta }\left( t-t_{0}\right) \right] }\right) , 
  \label{s3}
\end{equation}%
where we have denoted
\begin{equation}
\Delta _{0}=\left( H_{0}-\sqrt{\Lambda /3\eta }\right) \left( H_{0}+\sqrt{%
\Lambda /3\eta }\right) ^{-1},    \label{s4}
\end{equation}%
and we observe that $\Delta _{0}=0\rightarrow H=\sqrt{\Lambda /3\eta }$,
i.e., an usual early de Sitter phase (old inflation-like). If $\Delta
_{0}\neq 0$, the solution (40) is a reasonable one solution for $t<<t_{0}$
(very early times), i. e., $H\left( t<<t_{0}\right) \rightarrow $constant
and we note also that there is no singularity in $H\left( t\right) $ for $%
\Delta _{0}>0$. For completeness, with $\omega =1$ and $\eta <0$, the Eq.
(39) has the solution
\begin{equation}
H\left( t\right) =\sqrt{\Lambda /3\left\vert \eta \right\vert }\tan \left[
\arctan \left( \frac{H_{0}}{\sqrt{\Lambda /3\left\vert \eta \right\vert }}%
\right) -6\sqrt{\Lambda /3\left\vert \eta \right\vert }\left( t-t_{0}\right) %
\right] ,    \label{s5}
\end{equation}%
and this solution will be reasonable only if $t<<t_{0}$ (very early times
and, for instance, $0<\arctan \left( H_{0}/\sqrt{\Lambda /3\left\vert \eta
\right\vert }\right) +6\sqrt{\Lambda /3\left\vert \eta \right\vert }%
t_{0}<\pi /2\Longrightarrow H\left( t\right) >0$), i. e., $H\left(
t<<t_{0}\right) \rightarrow $constant. Thus, even the presence of $\omega =1$%
, is a cosmological constant the dominant component at early times. In other
words, even accepting the presence of stiff matter at early times, in HL
cosmology this is a little relevant fact (nevertheless, a fluid for which $%
p=\rho $ could play a relevant role at early times if we are thinking in to
build an holographic approach to cosmology, see [29] and [30]).
Finally, we compare the deceleration parameter as given in GR and HL
cosmology. This parameter is defined by $q=-\left( 1+\dot{H}/H^{2}\right) $
so that, in GR reads

\begin{equation}
q=\frac{1}{2}\left( 1+3\omega \right) -\frac{1}{2}\left( 1+\omega \right) 
\frac{\Lambda }{H^{2}},  \label{sl1}
\end{equation}%
and in HL cosmology

\begin{equation}
q=\frac{1}{2}\left( 1+3\omega \right) -\left( 1-\omega \right) \frac{\alpha
k^{3}}{\eta H^{2}a^{6}}-\frac{1}{2}\left( 1+\omega \right) \frac{\Lambda }{%
\eta H^{2}},  \label{sl2}
\end{equation}%
and for $\omega =1$ we have $q=2-\Lambda /H^{2}$ and $q=2-\Lambda /\eta
H^{2} $, respectively, and we note that when $\omega =1$, the curvature term 
$\sim \alpha k^{3}$ disappears.

\section{Final Remarks}

We have studied some aspects of the HL cosmology where the emphasis has been
put on discuss some cosmological solutions which are present too in GR, in
particular, an evolution driven by dust is the same in both theories (flat
case). A possible phantom stage has been discussed where we have found a
condition less restrictive over the $\omega $-parameter ($\omega <0$, not $%
\omega <-1$). At late times, the energy density exhibits a like dust
behavior nevertheless the $\omega $-parameter satisfies the inequality $%
\omega <-1/3$. If we do infinite the parameter which preserves the
diffeomorphism invariance in the present theory, a late de Sitter phase can
be obtained without consider the usual scheme $\omega =-1$. The combined
effect of a curvature-dependent term, which is stiff matter reminiscent and
dominant at early times, beside the usual stiff matter component ($\omega =1$%
, possible important role at early evolution) and a cosmological constant
has been discussed and solutions has been found which exhibit an early de
Sitter phase and this fact shows that in HL cosmology the role of stiff
matter is of little relevance.

\begin{acknowledgments}
This work has been supported by COMISI\'ON NACIONAL DE CIENCIAS Y TECNOLOG\'IA
through FONDECYT Grants 1110076 (JS and SL), 1090613 and 1110230 (JS). This work was also
partially supported by PUCV-VRIEA grant No. 037.492/2013  (SL) and PUCV grant No. 123.713/2012.
(JS).
\end{acknowledgments}

\end{document}